\documentclass[conference]{IEEEtran}

\makeatletter
\long\def\@makecaption#1#2{\ifx\@captype\@IEEEtablestring%
\footnotesize\begin{center}{\normalfont\footnotesize #1}\\
{\normalfont\footnotesize\scshape #2}\end{center}%
\@IEEEtablecaptionsepspace
\else
\@IEEEfigurecaptionsepspace
\setbox\@tempboxa\hbox{\normalfont\footnotesize {#1.}~~ #2}%
\ifdim \wd\@tempboxa >\hsize%
\setbox\@tempboxa\hbox{\normalfont\footnotesize {#1.}~~ }%
\parbox[t]{\hsize}{\normalfont\footnotesize \noindent\unhbox\@tempboxa#2}%
\else
\hbox to\hsize{\normalfont\footnotesize\hfil\box\@tempboxa\hfil}\fi\fi}
\makeatother

\ifCLASSINFOpdf
  \usepackage[pdftex]{graphicx}
  \usepackage{subfig}
\else
\fi

\usepackage{amsmath} 
\usepackage{algorithm}
\usepackage{algpseudocode}

\usepackage{amsthm}
\begin{document}
% \begin{spacing}{0.96}

\title{3D Multi-Drone-Cell Trajectory Design for Efficient IoT Data Collection
\vspace{-0.25cm}
}

\author{Weisen~Shi\IEEEauthorrefmark{1},
        Junling~Li\IEEEauthorrefmark{1},
        Nan~Cheng\IEEEauthorrefmark{1}\IEEEauthorrefmark{2},
        Feng~Lyu\IEEEauthorrefmark{1},
        Yanpeng~Dai\IEEEauthorrefmark{2},
        Haibo~Zhou\IEEEauthorrefmark{3},
        and~Xuemin~(Sherman)~Shen\IEEEauthorrefmark{1},~\IEEEmembership{Fellow,~IEEE}
% \thanks{W. Shi, J. Li, N. Cheng (cooresponding author), F. Lyu, and S. Shen are with the Department of Electrical and Computer Engineering, University of Waterloo, 200 University Avenue West, Waterloo, Ontario, Canada, N2L 3G1, (email: \{w46shi, j742li, n37cheng, f2lyu, sshen\}@uwaterloo.ca).}
% \thanks{N. Zhang is with the Department of Computing Science, Texas A\&M University-Corpus Christi, 6300 Ocean Dr., Corpus Christi, TX. 78412, (email: ning.zhang@tamucc.edu).}
% \thanks{S. Zhang is with the School of Computer Science and Engineering, Beihang University, Beijing, China, 100191, (email: zhangshan2007@gmail.com).}
% \thanks{H. Zhou is with the School of Electronic Science and Engineering, Nanjing University, Nanjing, China, 210023, (email: haibozhou@nju.edu.cn).}
% }

\IEEEauthorblockA{
\\\IEEEauthorrefmark{1}Department of Electrical and Computer Engineering, University of Waterloo, Waterloo, Ontario, Canada, N2L 3G1.
\\\IEEEauthorrefmark{2}State Key Laboratory of Integrated Service Networks, Xidian University, Xi’an, China, 710071.
\\\IEEEauthorrefmark{3}School of Electronic Science and Engineering, Nanjing University, Nanjing, China, 210093.
% \\\IEEEauthorrefmark{3}Department of Computing Science, Texas A\&M University-Corpus Christi, Corpus Christi, USA, 78412.
\\E-mail: \{w46shi, j742li, n5cheng, f2lyu, sshen\}@uwaterloo.ca. yp\_dai@stu.xidian.edu.cn. haibozhou@nju.edu.cn.}
\vspace{-0.5cm}
}

% make the title area
\maketitle

% As a general rule, do not put math, special symbols or citations in the abstract
\begin{abstract}
Drone cell (DC) is an emerging technique to offer flexible and cost-effective wireless connections to collect Internet-of-things (IoT) data in uncovered areas of terrestrial networks.
% Drone cell (DC) is an emerging technique to extend wireless connections for Internet-of-things (IoT) devices in spectrum-scared or uncovered areas of terrestrial base stations (BSs). 
The flying trajectory of DC significantly impacts the data collection performance. 
However, designing the trajectory is a challenging issue due to the complicated 3D mobility of DC, unique DC-to-ground (D2G) channel features, limited DC-to-BS (D2B) backhaul link quality, etc.
% However, how to design the trajectory of DCs to improve the IoT data collection performance is of great importance yet challenging due to the complicated 3D mobility of DC, unique DC-to-ground (D2G) channel features, limited DC-to-BS (D2B) backhaul link quality, etc.
In this paper, we propose a 3D DC trajectory design for the DC-assisted IoT data collection where multiple DCs periodically fly over IoT devices and relay the IoT data to the base stations (BSs). 
% In this paper, we investigate the scenario where multiple DCs periodically fly over IoT devices to relay their data transmission to the BS while the user (IoT devices)-to-DC (U2D) pathloss is minimized.
The trajectory design is formulated as a mixed integer non-linear programming (MINLP) problem to minimize the average user-to-DC (U2D) pathloss, considering the state-of-the-art practical D2G channel model. 
% According to the state-of-the-art D2G channel models, the trajectory design for multiple DCs is formulated to be a mixed integer non-linear programming (MINLP) problem with both user association and scheduling consideration.
We decouple the MINLP problem into multiple quasi-convex or integer linear programming (ILP) sub-problems, which optimizes the user association, user scheduling, horizontal trajectories and DC flying altitudes of DCs, respectively.
Then, a 3D multi-DC trajectory design algorithm is developed to solve the MINLP problem, in which the sub-problems are optimized iteratively through the block coordinate descent (BCD) method.
Compared with the static DC deployment, the proposed trajectory design can lower the average U2D pathloss by 10-15 dB, and reduce the standard deviation of U2D pathloss by 56\%, which indicates the improvements in both link quality and user fairness.
% In this paper, we propose the 3D multi-DC trajectory design algorithm to promote the data collection in IoT, in which multiple DCs periodically relay data uploading from IoT devices to the base station (BS).
\end{abstract}

% \begin{IEEEkeywords}
% Drone Communication, Drone Base Station, Trajectory Planning, DA-RAN, Block Coordinate Descent.
% \end{IEEEkeywords}
% no keywords
\IEEEpeerreviewmaketitle

\section{Introduction}
In the future Internet-of-things (IoT), the IoT devices are foreseen to be widely deployed to collect the data required by a myriad of applications. However, in certain areas where the terrestrial network coverage is unavailable due to signal blockage, spectrum scarcity or low IoT transmit power, collecting the IoT data becomes a challenging issue \cite{cheng2018air}.
% One challenging yet important issue faced by future Internet-of-things (IoT) is collecting data from massive uneven distributed devices in uncovered areas where the effective wireless links from devices to base station (BS) are prone to be blocked by obstacles \cite{cheng2018air}.
Although such an issue can be alleviated through densely deploying massive BSs or small-cells, the prohibitive capital expenditure (CapEx) and operating expenditure (OpEx) are unacceptable for IoT operators.
In addition, the IoT data traffic is spatio-temporally unbalanced and dynamic-distributed in dedicated areas \cite{tao2018locating}, which leads to frequent network congestion in the case of fix-deployed networks.
% Given the limited spectrum assigned for IoT communication, the temporal traffic congestion can frequently happen in various areas where additional communication resources are required.
% However, the current BSs are fixedly deployed based on the statistic of long-term traffic distribution of IoT devices, which lacks flexibility to adjust their deployments according to the traffic variations \cite{li2017joint}. 
To address the connection and flexibility shortages, the state-of-the-art drone communication technology is leveraged to support IoT data collections \cite{mozaffari2018tutorial}.

% Though the Base Station (BS) deployment is determined according to the long-term traffic statistic of users, such rigid RAN is still reluctant to ensure the QoS required by the users that are uneven and dynamically distributed in both spatial and temporal domains \cite{ye2018endmagazine}. 

% On the other hand, the terrestrial RAN of 5G networks are statically fixed in certain geographical locations with little flexibility to be re-deployed. 
% Though the Base Station (BS) deployment is determined according to the long-term traffic statistic of users, such rigid RAN is still reluctant to ensure the QoS required by the users that are uneven and dynamically distributed in both spatial and temporal domains \cite{ye2018endmagazine}. 
% To overcome the coverage ratio and flexibility challenges faced by current RAN, the emerging drone communication technology is leveraged by some pioneer researchers \cite{mozaffari2018tutorial}. 

Various field experiments have verified the potential of drone, i.e., unmanned aerial vehicle (UAV), to support communication services for terrestrial users.
Equipped with certain communication modules, the flying drone can perform as the drone cell (DC) which collects the data from IoT devices through the user (IoT device)-to-drone (U2D) wireless links.
Compared with the terrestrial BS, DC has distinct advantages:
\begin{itemize}
\item\emph{Line-of-sight (LoS) connection:}
User-to-BS (U2B) wireless links in terrestrial IoT are frequently blocked by the terrain or buildings, and such non-line-of-sight (NLoS) wireless links seriously degrade the U2B communications \cite{liu2018spaceairground}.
In contrast, the flying DCs are capable of adjusting 3D positions, which can ensure higher probability to connect terrestrial IoT devices via the highly reliable LoS links \cite{mozaffari2018tutorial}.
\item\emph{Dynamic deployment:}
Compared with terrestrial BSs built on fixed locations, DCs can change their deployments according to the spatio-temporal traffic variations, and assigned to different controllers or IoT devices on demands \cite{xue2018maximization}. 
\item\emph{Fully-controlled mobility:}
% Compared with the connected vehicles in VANET whose mobility is not determined by network providers, the hovering positions or flying trajectory of any DC are fully controlled by the operators, which enables the dynamic deployment feature \cite{zhang2018air}.
The flying traces and behaviors of any DC are fully controlled by the network providers, which empowers DCs with the dynamic deployment feature and facilitates inter-DC collaborations \cite{zhang2018air}.
\end{itemize}

Leveraging the advantages of DC, how to control the 3D placements of DCs to improve the network performance is of great importance yet very challenging due to the complicated 3D mobility of DC and dynamic DC-to-ground (D2G) link quality.
In the literature, there have been several studies on the 3D placements of single/multiple DCs to support terrestrial IoT, which can be classified as two categories, i.e., static DC deployment and DC trajectory design. 
The static DC deployment investigates the optimal hovering positions of DCs to maximize the communication performance of associated users.
Bor-Yaliniz \emph{et al.} proposed a bisection search based algorithm to deploy single DC \cite{bor2016efficient}. 
Kalantari \emph{et al.} optimized the multiple DCs static deployment through the swarm intelligence based algorithm \cite{kalantari2016number} \cite{kalantari2017backhaul}. 
% In \cite{zhang2017spectrum}, Zhang \emph{et al.}. studied the spectrum sharing of Drone-Small-Cells (DSCs) network, and calculated the optimal DC density that maximizes the network throughput without violating the efficiency requirements.
% Alzenad \emph{et al.}. designed an optimal DC placement algorithm to maximize the number of effectively covered users while satisfying the minimum energy consumption requirement \cite{alzenad20173}.
% As the extension work of \cite{kalantari2016number}, Kalantari \emph{et al.} further optimize the 
% Kumar \emph{et al.}. propose the backhaul and delay aware DC deployment algorithm to determine the optimal deployment height \cite{kumar2018backhaul}.
% In \cite{shah2017association} and \cite{shah2017distributed}, a swarm of DCs were regarded as backhaul/fronthaul hubs for small-cells via free-space-optics /mmWave links by Shah \emph{et al.}., in which the DC deployment and association is jointly optimized through heuristic methods.
To maximize the IoT information collection gain, Mozaffari \emph{et al.} leveraged the clustering based method to design optimal hovering positions for multiple DCs \cite{mozaffari2017mobile}.
% In \cite{ghanavi2018efficient}, even the $Q$-learning approach was applied by Ghanavi \emph{et al.}. to solve the static DC deployment problem too.
% Yang \emph{et al.}. designed a holistic framework using DCs to assist 5G networks in flash crowd traffic scenarios, and proposed a ``first-selfish and second-share'' method to calculate the deployment of the DCs \cite{yang2017proactive}.
However, the static deployment can hardly guarantee the fairness among users.
Particularly, the IoT devices located at the edge of the DC's radio coverage can suffer severer pathloss compared with devices located at the center of the radio coverage, which leads to unbalanced communication performance.
% However, there is an inherent weakness of the static deployment scheme, i.e., lacking of the fairness guarantee. 
% , while the static deployment scheme fails to guarantee the performance fairness for devices unevenly located in the environment.

\begin{figure}[htpb]
%   \vspace{-0.1cm}
  \centering
  \includegraphics[width=0.48\textwidth]{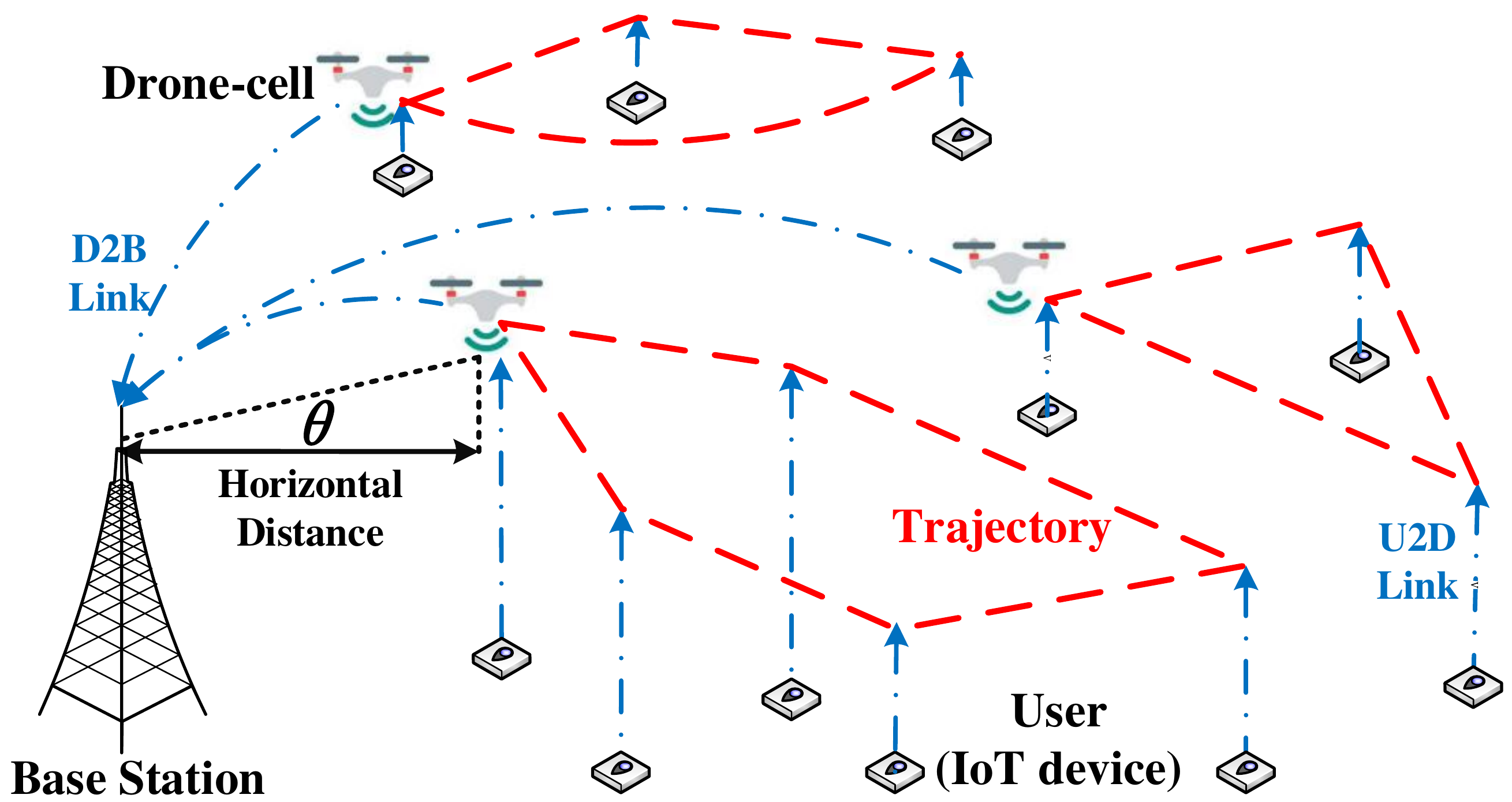}
  \caption{DC enabled IoT data collection.}
  \label{DARAN}
  \vspace{-0.4cm}
\end{figure}
% However, the D2B backhaul channel models used in those works are either U2D pathloss model or traditional terrestrial channel models. 
% In this paper, we further implement the specific D2B channel model derived in \cite{al2017modeling} to highlight the D2B channel features.
% According to the state-of-the-art U2D channel model \cite{al2014optimal}, the IoT devices located at the edge and center of the DC's radio coverage suffer different levels of pathloss. 
To promote the fairness among IoT devices, DC trajectory design schemes are proposed in some recent works, in which DCs periodically fly over associated users via optimized trajectories, and communicate with dedicated users in scheduled slots [12]-[16].
% For instance, Li \emph{et al.}. investigate the sensor to BS packets transmission relayed by multiple DCs following pre-determined DC trajectories \cite{li2016energy}.
Following pre-determined DC trajectories, Li \emph{et al.} investigated the packets transmission relayed by multiple DCs from the sensor to BS \cite{li2016energy}.
% Mozaffari \emph{et al.} study the constrained DC trajectory for DC enabled wireless networks \cite{mozaffari2016unmanned}. 
Zeng \emph{et al.} proposed a general framework for trajectory and communication optimization of a energy-efficient point-to-point UAV-ground communication system \cite{zeng2017energy}. 
%The trajectory designs in \cite{zeng2016throughput} and \cite{zeng2017energy} can be considered as a generalization of the DCs trajectory planning problem, subject to practical constraints on the DC’s mobility, such as its initial/final locations, maximum speed and acceleration, etc. 
% Lyu \emph{et al.}. proposed a novel cyclical multiple access scheme for the UAV-enabled multi-user communication networks \cite{lyu2016cyclical}.
In \cite{wu2017joint}, Wu \emph{et al.}. formulated the non-convex DC trajectory design problem in which the user association, DC horizontal trajectory design and DC transmitting power control are jointly optimized;
the delay constraint is further considered in the extension work focusing on delay-sensitive services \cite{wu2018common}.
However, there are some strong assumptions and model simplifications in current DC trajectory design studies. 
% However, there are still some limitations in current DC trajectory design studies in terms of model simplification and strong assumptions. 
For instance, in current works, the flying altitudes of all DCs are usually set to be a constant value to simplify the optimization process (i.e., only 2D trajectory is optimized), and the adopted D2G channel model is the classic Friis model which fails to characterize the distinct features of DCs \cite{wu2017joint} \cite{wu2018common}. 
Besides, most existing trajectory design works idealize or ignore the drone DC-to-BS (D2B) link quality constraints, which however is essential for drone-assisted networks. 

% Both \cite{wu2017joint} and \cite{wu2018common} set the foundation models for DC trajectory design.
% While the flying altitudes of all DCs are set as a constant to simplify the optimization, the D2G channel model is the classic Friis model without involving DC features.
% Besides, most existing trajectory design works idealize or ignore the D2B link quality constraints.
% However, limited by the insufficient computing capacity embedded on DCs and the global data analysis requirements for most IoT services, the essential task of DC in IoT data collection is relaying data transmission from IoT devices to the BS rather than processing data.
% Since that, it is inevitable to ignore the D2B link quality constraints in the DC enabled IoT data collection.
To address the aforementioned issues, in this paper, we study the multiple DCs 3D trajectory design to facilitate the IoT data collection.
We leverage the state-of-the-art U2D \cite{al2014optimal} and D2B \cite{al2017modeling} pathloss models, and the DC altitudes are taken as optimization variables to further improve the network performance.
We formulate the 3D multi-DC trajectory design as a mixed integer non-linear programming (MINLP) problem, in which the summation of average U2D pathloss is minimized under the constraints of D2B link qualities, user fairness and horizontal/vertical speed limitations.
Given the constant transmit power, bandwidth and interference-level of IoT devices, the lower average U2D pathloss indicates higher network throughput. 
% Leveraging the state-of-the-art U2D \cite{al2014optimal} and D2B \cite{al2017modeling} pathloss models, a mixed integer non-linear programming (MINLP) problem is formulated to describe the 3D multi-DC trajectory design problem, in which the summation of average U2D pathloss is minimized under the constraints of D2B link qualities, user fairness and horizontal/vertical speed limitations.
As the formulated MINLP problem is very difficult to be solved directly, we decouple the original MINLP problem into multiple quasi-convex or linear sub-problems.
Each sub-problem optimizes the user association, user scheduling, DC horizontal trajectories and flying altitudes, respectively.
By adopting the block coordinate descent (BCD) method, we then propose a 3D multi-DC trajectory design algorithm in which those sub-problems are iteratively optimized to achieve the optimal result of the MINLP problem.
According to the simulation results, our algorithm can reduce the average D2U pathloss by $10-15~\mathrm{dB}$, and lower the standard deviation by 56\% when compared with the static DC deployment scheme.

% Compared with the static DC deployment result based on particle intelligent heuristic \cite{kalantari2016number} , the average U2D pathloss and U2D pathloss standard deviation can be reduced by $15-20 \mathrm{dB}$, and more than $40\%$ respectively when applying proposed trajectory design algorithm.

The remainder of this paper is organized as follows. 
In Section II, we introduce the system model. 
Then, we formulate the trajectory design problem for multiple DCs in Section III. 
In Section IV, we propose the 3D multi-DC trajectory design algorithm to solve the problem.
Numerical results are shown in Section V, and the conclusion are given in Section VI.

\section{System Model}
The overview of DC enabled IoT data collection is shown in Fig. \ref{DARAN}, where multiple DCs are released to relay data uploading from IoT devices to one BS.
We define the DC users as the IoT devices whose reliable U2B links are compromised due to blockage or resource shortage.
The BS periodically detects the channel-state-information (CSI) of each U2B pair and assign compromised IoT devices into the DC user set $\mathcal{U}$. 
Given $\mathcal{U}$ and the available DCs set $\mathcal{D}$, the BS implements the trajectory design algorithm to determine the optimal trajectory for each DC.
% We assume DC users are uniformly distributed over the BS's radio coverage area.
% Without loss of generality, all 
DC users are considered as identical devices which transmit data to the BS with identical data collection period, bandwidth, and transmit power.
Since the proposed trajectory design algorithm can schedule the transmit time of all users associated to one DC to prevent transmit-interval overlapping, and we assume each DC can distinguish its associated users' signal efficiently from others, the inter-user interference issue is prevented in our system.
The cardinalities $|\mathcal{U}|$ and $|\mathcal{D}|$ represent the number of DC users and available DCs, respectively.

\subsection{U2D and D2B Channel Models}
Both U2D and D2B wireless links are modelled based on the recent D2G channel research \cite{al2014optimal} \cite{al2017modeling}.
According to \cite{al2014optimal}, we can express the U2D pathloss as
\begin{equation}
\begin{aligned}
{PL(r_{\mathrm{UD}},h)} &= 10\log(\frac{16 {\pi}^2 {f_{c}}^2 (h^2+r_{\mathrm{UD}}^2)}{c^2})\\
&+ P_{\mathrm{LoS}}\eta_{\mathrm{LoS}} + (1 - P_{\mathrm{LoS}})\eta_{\mathrm{NLoS}}
\end{aligned}
\label{pathlossEq}
\end{equation}
where $r_{\mathrm{UD}}$ is the U2D horizontal distance, and $h$ is the DC flying altitude.
$f_{c}$ and $c$ are carrier frequency and speed of light in Hz and m/s, respectively. 
$\eta_{\mathrm{LoS}}$ and $\eta_{NLoS}$ are environment-dependent LoS and NLoS pathloss offsets, respectively.
$P_{\mathrm{LoS}}$ is the U2D LoS probability, which is expressed as
\begin{equation}
{P_{\mathrm{LoS}}(r_{\mathrm{UD}},h)} = \frac{1}{1 + a e^{(-b(\arctan(\frac{h}{r_{\mathrm{UD}}}) - a))}}
\label{P_LoS}
\end{equation}
where $a$ and $b$ are environment-dependent parameters. 
To prevent the interference to U2B communications, as well as to provide additional spectrum resource for DC users, $f_{c}$ is expected to be different from the traditional IoT band, such as Wi-Fi or NB-IoT \cite{soussi2018nbiot} bands.
% Currently, commercial drone products leverage WiFi band to support U2D communication \cite{bisio2018unauthorized}.
% The TV White Space (TVWS) band is also considered by some researchers \cite{cheng2018air}.
% In our work, the $2.4\mathrm{GHz}$ WiFi band is chosen to carry U2D communications.
% D2B links are designed to provide high-reliable data transmission between DCs and their corresponding BS.
The average D2B pathloss is calculated by \cite{al2017modeling}
\begin{equation}
\begin{aligned}
{PL(r_{\mathrm{DB}},\theta)} = 10\alpha\log(r_{\mathrm{DB}}) + A(\theta - \theta_{0})\mathrm{e}^{(\frac{\theta_{0}-\theta}{B})}+\eta_{0}
\end{aligned}
\label{pathlossEqD2B}
\end{equation}
where $r_{\mathrm{DB}}$ and $\theta$ denote the D2B horizontal distance and the DC to BS antenna vertical angle, respectively. 
$\alpha$, $A$, $\theta_{0}$, $B$, and $\eta_{0}$ are all environment-dependent parameters.
% For the suburban scenario investigated by \cite{al2017modeling}, the values of parameter list $(\alpha,A,\theta_{0},B,\eta_{0})$ are $(3.04,-23.29,-3.61,4.14,20.7)$. 
Since the 850 MHz band is used for LTE transmissions \cite{al2017modeling}, (\ref{pathlossEqD2B}) contains no parameter representing carrier frequency.
% In this work we set the D2B carrier frequency as $850\mathrm{MHz}$ to keep alignment with the D2B pathloss model.
% Though the reliability of D2B links can be guaranteed according to the D2B channel model \cite{al2017modeling}, the D2B link capacity remains an issue since each D2B link has to support all users served by the DC.
% The millimeter wave (mmWave) technology, which can provide up to 20Gbps data-rate transmission, is regarded as an promising solution to extend D2B communication capacity \cite{zhang2017is}.
% On the other hand, the mmWAVE technology can reach its best performance in LoS environment \cite{thornburg2016performance}, which is naturally featured by D2B links. 
% The two pathloss models are suitable for both up/down link communications. 
% Without loss of generality, in this work we focus on the downlink contents distribution scenario where multiple DCs deliver contents to users in user.

\subsection{DC Trajectory Model}
For one DC $d \in \mathcal{D}$, we consider that it serves the associated user set $\mathcal{U}_d \subseteq \mathcal{U}$ through a time division multiple access (TDMA) mode.
Within one trajectory period $T$, DC $d$ flies over all its associated users and sequentially severs them according to the scheduling result.
$T$ is set to the same value as the data collection period of users, in order that each user can transmit the collected data once within one period.  
% To coordinate with the user, the flying trajectory period $T$ is set as same as the U2B data transmission period. 
By discretizing $T$ into $N$ equal-time slots, trajectory of DC $d$ within one period can be modeled as an $N$-length sequence composed by three-dimensional vectors: 
\begin{equation}
\begin{aligned}
\mathbf{G}_d[n] = [x_d[n], y_d[n], h_d[n]], \quad n \in \mathcal{N}
\end{aligned}
\label{trajectoryModel}
\end{equation}
where $x_d[n]$, $y_d[n]$ and $h_d[n]$ denote the 3D coordinates of DC $d$ at slot $n$.
$\mathcal{N}$ is the set of slots in one period.
In this work, we define $a_{u,d} = 1$ when user $u$ is associated to DC $d$ for $d \in \mathcal{D}, u \in \mathcal{U}$, otherwise $a_{u,d} = 0$.
The U2D communication scheduling is denoted by the binary variable $k_{u,d}[n]$ for $\forall n \in \mathcal{N}, d \in \mathcal{D}, u \in \mathcal{U}$.
If user $u \in \mathcal{U}$ is severed by DC $d$ in slot $n$, $k_{u,d}[n]$ is set as $1$; otherwise, $k_{u,d}[n] = 0$.

Several trajectory constraints are considered in our work: 
1) The trajectory of each DC has to be a 3D closed curve since the DC must return to the start point by the end of each period $T$ for a new period.
2) Since the power consumption models for horizontal and vertical moving of DC are different, it is reasonable to set the maximum horizontal speed $V_\mathrm{max}$ and maximal vertical speed $H_\mathrm{max}$, respectively.
% \cite{tseng2017autonomous}
3) In any slot $n$, one DC $d$ can serve at most one user $u \in \mathcal{U}_d$; in all slots, one user $u$ can only be associated to one DC.
4) The slots amount scheduled to one $u$ within one period cannot be smaller than a pre-defined threshold $S_\mathrm{min}$, which indicates the minimal time to complete each U2B data transmission.
% 5) To prevent frequent user switching in TDMA mode, all slots scheduled to one $u$ are consecutive.

\section{Problem Formulation}
% In this section we formulate the 3D multi-DC trajectory planning problem based on the aforementioned system model.
% The DC-user association is denoted by the binary variable $k_{u,d}[n]$ for $\forall n,d,u$.
% If user $u \in \mathcal{U}$ is severed by DC $d$ in slot $n$, $k_{u,d}[n]$ is set as $1$; otherwise, $k_{u,d}[n] = 0$.
% For each DC with pre-defined trajectory planning and user association results, a U2D communication scheduling scheme is designed to allocate each TDMA slot to the corresponding user, and guarantee the fairness among all user in $\mathcal{U}_d$.

% Based on  (\ref{trajectoryModel}), the 3D distance from the DC $d$ to user $u$ in time slot $n$ can be calculated as:
% \begin{equation}
% \begin{aligned}
% m_{u,d}[n] & = \sqrt{h_d[n]^2 + {\| \mathbf{R}_d[n] - \mathbf{R}_u \|}^2}
% %= \sqrt{h_d[n]^2 + r_{u,d}[n]^2}
% \end{aligned}
% \label{trajectoryDist}
% \end{equation}
We define $\mathbf{R}_u = [x_u, y_u]$ as the coordinate of $u$, $\mathbf{R}_d[n] = [x_d[n], y_d[n]]$ as the 2D projection of DC $d$ on X-Y plane, and $r_{u,d}[n] = {\| \mathbf{R}_d[n] - \mathbf{R}_u \|}$ as the horizontal distance between $d$ and $u$.
% Define the elevation angle between DC $d$ and user $u$ as $\theta_{d,u}[n] = \arctan(h_d[n]/{\| \mathbf{R}_d[n] - \mathbf{R}_u \|})$ in degree, and 
Define the elevation angle between DC $d$ and BS antenna as $\theta_{d,B}[n] = \arctan(h_d[n]/{\|\mathbf{R}_d[n]\|})$ in degree.
$\|\bullet\|$ is the Euclidean norm.

By substituting $r_{u,d}[n]$ and $\theta_{d,B}[n]$ into (\ref{pathlossEq}) and (\ref{pathlossEqD2B}) respectively, we can calculate the U2D pathloss between DC $d$ and user $u$ in slot $n$ by
\begin{equation}
\begin{aligned}
L_{u,d}[n] & = 20\log(\frac{4 \pi f_{c} \sqrt{r_{u,d}[n]^2 + h_{u,d}[n]^2}}{c}) + \eta_{\mathrm{NLoS}} \\
           & + P_{\mathrm{LoS}}(r_{u,d}[n], h_{u,d}[n])(\eta_{\mathrm{LoS}} - \eta_{\mathrm{NLoS}})
\end{aligned}
\label{U2DpathlossSub}
\end{equation}
as well as the D2B pathloss between the BS and DC $d$ by
\begin{equation}
\begin{aligned}
L_{d,B}[n] & = 10\alpha\log({\|\mathbf{R}_d[n]\|}) \\
           & + A(\theta_{d,B}[n] - \theta_{0})\mathrm{e}^{(\frac{\theta_{0}-\theta_{d,B}[n]}{B})}+\eta_{0}.
\end{aligned}
\label{D2BpathlossSub}
\end{equation}
% where $\theta_{d,B}[n] = \arctan(h_d[n]/{\|\mathbf{R}_d[n]\|})$ in degree.

Assuming all users maintain their transmission power $P_{tu}$ in each period, the effectiveness of the data collection for each U2D link is determined by the U2D pathloss.
Therefore, we aim to minimize the total pathloss suffered by each DC in the trajectory design.
Define the matrices $\mathbf{U} = \{a_{u,d}, \forall d,u\}$, $\mathbf{K} = \{k_{u,d}[n], \forall d,u,n\}$ and $\mathbf{G} = \{\mathbf{G}_d[n], \forall d,n\}$.
% To minimize the total pathloss suffered by each DC during one period $T$, t
The 3D multi-DC trajectory design problem can be formulated as
\begin{subequations}
\begin{align}
& \min_{\mathbf{U},\mathbf{K},\mathbf{G}} 
\quad \sum\limits_{d=1}^{|\mathcal{D}|}\sum\limits_{u=1}^{|\mathcal{U}|}{a_{u,d}}(\sum\limits_{n=1}^{N}k_{u,d}[n]L_{u,d}[n]) \\
      & s.t. \quad \sum\nolimits_{u=1}^{|\mathcal{U}|} a_{u,d} \le N_u, \quad \forall d, \\
      & \quad \quad \sum\nolimits_{d=1}^{|\mathcal{D}|} a_{u,d} = 1, \quad \forall u, \\
      & \quad \quad \sum\nolimits_{u=1}^{|\mathcal{U}|} k_{u,d}[n] = 1, \quad \forall d,n, \\
      & \quad \quad \sum\nolimits_{d=1}^{|\mathcal{D}|} k_{u,d}[n] = 1, \quad \forall u,n, \\
      & \quad \quad \sum\nolimits_{n=1}^{N} k_{u,d}[n] \ge S_\mathrm{min}, \quad \forall u,d, \\
      & \quad \quad a_{u,d}, k_{u,d}[n] \in \{0,1\}, \quad \forall d,u,n, \\
    %   & \quad \quad \max(\sum\limits_{o=1}^{\frac{N}{|\mathcal{U}_d|}}k_{u,d}[{(n+o)}\bmod{N})]) = \frac{N}{|\mathcal{U}_d|}, \quad \forall u \in \mathcal{U}_d,d,n, \\
      & \quad \quad \mathbf{G}_d[1] = \mathbf{G}_d[N+1], \quad \forall d,\\
	  & \quad \quad {\| \mathbf{R}_d[n+1] - \mathbf{R}_d[n] \|} \le {V_\mathrm{max}\delta_{t}}, \quad \forall d,n, \\
	  & \quad \quad {|h_d[n+1] - h_d[n]|} \le H_\mathrm{max}\delta_{t}, \quad \forall d,n, \\
% 	  & \quad \quad {\|\mathbf{G}_i[n] - \mathbf{G}_j[n]\|} \ge Z_\mathrm{min}, \quad \forall n,i,j \ne i, \\
	  & \quad \quad L_{d,B}[n] \le \mathrm{L}_{\mathrm{DB}}, \quad \forall d,u,n.
\end{align}
\label{optimal}
\end{subequations}
% In (\ref{optimal}), $S_\mathrm{min}$ denotes the minimum number of slots allocated to each $u$ in one period.
In (\ref{optimal}), $N_u$ is the maximum number of users for one $\mathcal{U}_d$, and $\mathrm{L}_{\mathrm{DB}}$ is the D2B pathloss threshold.
(\ref{optimal}c)-(\ref{optimal}g) correspond to user association constraint 3) and 4) in section II. 
(\ref{optimal}h)-(\ref{optimal}j) represent the DC trajectory constraints 1) and 2) in section II. 
(\ref{optimal}k) is the D2B pathloss constraint.

Due to the quadratic and exponential terms in (\ref{optimal}a) and (\ref{optimal}k), and the binary variable $k_{u,d}[n]$ and $a_{u,d}$, the 3D multi-DC trajectory design problem turns to be an MINLP problem \cite{bor2016efficient}.
Besides, the objective function (\ref{optimal}a) and constraint (\ref{optimal}n) are both non-convex for DC trajectory $\mathbf{G}$, which increases the difficulty to solve the problem.

\section{3D Multi-DC Trajectory Design Algorithm}
The non-convex problem (\ref{optimal}) can be transformed into solvable forms (e.g. quasi-convex or LP) by assuming parts of the decision variables as constants.
Then, we can decouple the MINLP problem into multiple sub-problems that are solvable for parts of the decision variables, and iteratively solve them by applying the BCD method.
Specifically, we divide the decision variable in problem (\ref{optimal}) into four blocks (i.e. $\mathbf{U}$, $\mathbf{K}$, $\mathbf{R} = \{r_{u,d}[n], \forall d,u,n\}$ and $\mathbf{H} = \{h_{u,d}[n], \forall d,u,n\}$), and formulate following sub-problems.

Given the pre-defined trajectories of multiple DCs (constant $\mathbf{K}$, $\mathbf{R}$ and $\mathbf{H}$), the user association sub-problem can be written as an integer linear programming (ILP) problem:
\begin{equation}
\begin{aligned}
& \min_{\mathbf{U}} 
\quad \sum\limits_{d=1}^{|\mathcal{D}|}\sum\limits_{u=1}^{|\mathcal{U}|}{a_{u,d}}(\sum\limits_{n=1}^{N}k_{u,d}[n]L_{u,d}[n]) \\
& s.t. \quad (7b), (7c), \quad a_{u,d} \in \{0,1\} \quad \forall d,u.
\end{aligned}
\label{subpAssociation}
\end{equation}
Problem (\ref{subpAssociation}) can be efficiently solved through the branch and bound method, which is well supported by Gurobi solvers. 

Based on the optimized $\mathbf{U}$ and pre-defined trajectory, the U2D communication scheduling sub-problem is also transformed to an ILP problem:
\begin{equation}
\begin{aligned}
& \min_{\mathbf{K}} 
\quad \sum\limits_{d=1}^{|\mathcal{D}|}\sum\limits_{u=1}^{|\mathcal{U}|}{a_{u,d}}(\sum\limits_{n=1}^{N}k_{u,d}[n]L_{u,d}[n]) \\
& s.t. \quad (7d), (7e), (7f), \quad k_{u,d}[n] \in \{0,1\} \quad \forall d,u,n.
\end{aligned}
\label{subpScheduling}
\end{equation}
Same as problem (\ref{subpAssociation}), the Gurobi optimizer can efficiently solve problem (\ref{subpScheduling}). 

According to (\ref{D2BpathlossSub}), for given $\mathbf{U}$, $\mathbf{K}$ and $\mathbf{H}$, (\ref{optimal}a) is still non-convex for $\mathbf{R}$.
However, by further decoupling $\mathbf{R}$ into element optimization variable $r_{u,d}[n]$, (\ref{optimal}a) is both quasi-convex and non-decreasing function to U2D horizontal distance $r_{u,d}[n], \forall d,u,n$, when other $r_{\bar{u},\bar{d}}[\bar{n}], \forall \bar{d} \neq d, \bar{u} \neq u, \bar{n} \neq n$ keep constants.
Since (\ref{optimal}a) can reach its minimal value when $r_{u,d}[n] = \min{r_{u,d}[n]}$ for any $r_{u,d}[n]$ within its available range, the objective of minimizing (\ref{optimal}a) equals minimizing $r_{u,d}[n]^2$, which is a quadratic convex function for $\mathbf{R}_d[n]$.
On the other hand, \cite{shi2018multiple} proves that the available range of $\mathbf{R}_d[n]$ for constraint (\ref{optimal}k) forms a convex set in X-Y plane. 
Therefore, the sub-problem to find optimal $\mathbf{R}_d[n]$ can be formulated as
\begin{equation}
\begin{aligned}
& \min_{\mathbf{R}_d[n]} 
\quad a_{u,d}[n]k_{u,d}[n]L_{u,d}[n] + \\ 
& \quad \quad \quad \sum\limits_{d=1}^{|\mathcal{D}|}\sum\limits_{u=1}^{|\mathcal{U}|}{a_{u,d}}(\sum\limits_{\bar{n}=1, \bar{n} \neq n}^{N}k_{u,d}[\bar{n}]L_{u,d}[\bar{n}]) \\
& s.t. \quad (7i), (7k), \quad \mathbf{R}_d[1] = \mathbf{R}_d[N+1], \quad \forall d
\end{aligned}
\label{subpHorTraj}
\end{equation}
whose optimal $\mathbf{R}_d[n]$ can be calculated by solving the following quadratic convex optimization problem:
\begin{equation}
\begin{aligned}
& \min_{\mathbf{R}_d[n]} 
\quad \| \mathbf{R}_d[n] - \mathbf{R}_u \|^2 \\
& s.t. \quad (7i), (7k), \quad \mathbf{R}_d[1] = \mathbf{R}_d[N+1], \quad \forall d.
\end{aligned}
\label{subpHorTrajEql}
\end{equation}
% Gurobi can efficiently solve the convex optimization problem (\ref{subpHorTrajEql}).

For any given $\mathbf{U}$, $\mathbf{K}$ and $\mathbf{R}$, the sub-problem optimizing $\mathbf{H}$ is also non-convex to $\mathbf{H}$ due to the $\log$-form and a $e^{1/\arctan}$-form terms.
Same as preceding decoupling process of $\mathbf{R}$, $\mathbf{H}$ can also be decoupled into element optimization variable $h_{u,d}[n]$, and the sub-problem to optimize $h_{u,d}[n]$ is 
\begin{equation}
\begin{aligned}
& \min_{h_{u,d}[n]} 
\quad a_{u,d}[n]k_{u,d}[n]L_{u,d}[n] + \\ 
& \quad \quad \quad \sum\limits_{d=1}^{|\mathcal{D}|}\sum\limits_{u=1}^{|\mathcal{U}|}{a_{u,d}}(\sum\limits_{\bar{n}=1, \bar{n} \neq n}^{N}k_{u,d}[\bar{n}]L_{u,d}[\bar{n}]) \\
& s.t. (7j), (7k), \quad h_d[1] = h_d[N+1] \quad \forall d.
\end{aligned}
\label{subpHeight}
\end{equation}
%  (\ref{optimal}a) can be proved quasi-convex to any $h_{u,d}[n], \forall d,u,n$ with other $h_{\bar{u},\bar{d}}[\bar{n}], \forall \bar{d} \neq d, \bar{u} \neq u, \bar{n} \neq n$ constant.
Problem (\ref{subpHeight}) can only be proved quasi-convex to $h_{u,d}[n], \forall d,u,n$ when other $h_{\bar{u},\bar{d}}[\bar{n}], \forall \bar{d} \neq d, \bar{u} \neq u, \bar{n} \neq n$ keep constant. 
However, since $L_{u,d}[n]$ is the summation of one non-increasing function of $h_{u,d}[n]$ and one non-decreasing function of $h_{u,d}[n]$, we can argue that problem (\ref{subpHeight})'s objective function has only one global minimum.
% To optimize $\mathbf{H}$, we transform  (\ref{U2DpathlossSub}) as the function of $r_{u,d}[n]$ and $\theta_{u,d}[n] = \arctan(h_d[n]/r_{u,d}[n])$:
% \begin{equation}
% \begin{aligned}
% L_{u,d}[n] & = 20\log(r_{u,d}[n]\sec(\theta_{u,d}[n])) + 20\log(\frac{4{\pi}f_{c}}{c}) \\
% & + \frac{\eta_{\mathrm{LoS}} - \eta_{\mathrm{NLoS}}}{1 + a\exp(-b(\theta_{u,d}[n] - a))} + \eta_{\mathrm{NLoS}}.
% \end{aligned}
% \label{U2DpathlossThetaSub}
% \end{equation}
% Since $r_{u,d}[n]$ is pre-defined constant, optimizing $\mathbf{H}$ is equivalent to optimizing every $\theta_{u,d}[n]$.
% Because  (\ref{U2DpathlossThetaSub}) is composed of an increasing function of $\theta_{u,d}[n]$ and a decreasing function of $\theta_{u,d}[n]$, we can argue that the $L_{u,d}[n]$ curve has only one global minimum. 
To achieve the optimal $h_{u,d}[n]_\mathrm{opt}$ which minimizes problem (\ref{subpHeight}), we can leverage the Newton-Raphson method by iteratively calculating the following function:
\begin{equation}
\begin{aligned}
h_{u,d}[n]_{i+1} = h_{u,d}[n]_{i} - \frac{{L_{u,d}[n]}(h_{u,d}[n]_{i})}{{L_{u,d}[n]}^{\prime}(h_{u,d}[n]_{i})}
\end{aligned}
\label{FirstDerivationTheta}
\end{equation}
where the iteration stops when $h_{u,d}[n]_{i+1} - h_{u,d}[n]_{i} \le \epsilon$ and $h_{u,d}[n]_\mathrm{opt} = h_{u,d}[n]_{i+1}$.

The sub-problems of problem (\ref{optimal}) can all be optimized respectively with other optimization variables keeping constant.
Therefore, problem (\ref{optimal}) can be solved through iteratively optimizing those sub-problems until the results converge, which yields to the classic BCD method.
Based on the BCD method, we propose the 3D multi-DC trajectory algorithm as shown in Algorithm \ref{algBCD}.
$\mathbf{U}_{t}$, $\mathbf{K}_{t}$, $\mathbf{G}_{t}$ denote the multiple DCs' user association, U2D communication scheduling and DC trajectories calculated after iteration $t$, respectively. 
Since the global optimal results of each sub-problem can be achieved accurately, the proposed algorithm ensures convergence \cite{wu2017joint} \cite{bertsekas1999nonlinear}.
\begin{algorithm}[htbp] 
\caption{3D multi-DC trajectory design algorithm} 
\begin{algorithmic}[1]
\State Initiate DC set $\mathcal{D}$ and their initial trajectory $\mathbf{G}_{0}$ composed by $\mathbf{R}_{}$ and $\mathbf{H}_{0}$. 
\State Initiate initial U2D communication scheduling $\mathbf{K}_{0}$.
\State $t = 1$, $\Delta{G}$ = $\infty$.
\While{$\Delta{G} \ge \mathbf{\epsilon}$}
  \State Solve (\ref{subpAssociation}) to obtain $\mathbf{U}_{t}$ with $\mathbf{K}_{t-1}$, $\mathbf{R}_{t-1}$ and $\mathbf{H}_{t-1}$.
  \State Solve (\ref{subpScheduling}) to obtain $\mathbf{K}_{t}$ with $\mathbf{U}_{t}$, $\mathbf{R}_{t-1}$ and $\mathbf{H}_{t-1}$.
  \For{$n \in \mathcal{N}, d \in \mathcal{D}$}
    \State Solve (\ref{subpHorTraj}) for $\mathbf{R}_{d}[n]_\mathrm{opt}$ with $\mathbf{U}_{t}$, $\mathbf{K}_{t}$ and $\mathbf{H}_{t-1}$.
    \State Update $\mathbf{R}_{t}$ with $\mathbf{R}_{d}[n]_\mathrm{opt}$.
  \EndFor
  \For{$n \in \mathcal{N}, d \in \mathcal{D}$}
    \State Solve (\ref{subpHeight}) for $h_{u,d}[n]_\mathrm{opt}$ with $\mathbf{U}_{t}$, $\mathbf{K}_{t}$ and $\mathbf{R}_{t}$.
    \State Update $\mathbf{H}_{t}$ with $h_{u,d}[n]_\mathrm{opt}$.
  \EndFor
  \State Update $\mathbf{G}_{t}$ with $\mathbf{R}_{t}$ and $\mathbf{H}_{t}$.
  \State $t = t + 1$.
  \State $\Delta{G} = \mathbf{G}_{t} - \mathbf{G}_{t-1}$.
\EndWhile
\end{algorithmic}
\label{algBCD} 
\end{algorithm}

\begin{figure*}[htbp]
  \centering
%   \subfigure[]{
  \subfloat[3D view]{\includegraphics[width=0.33\textwidth]{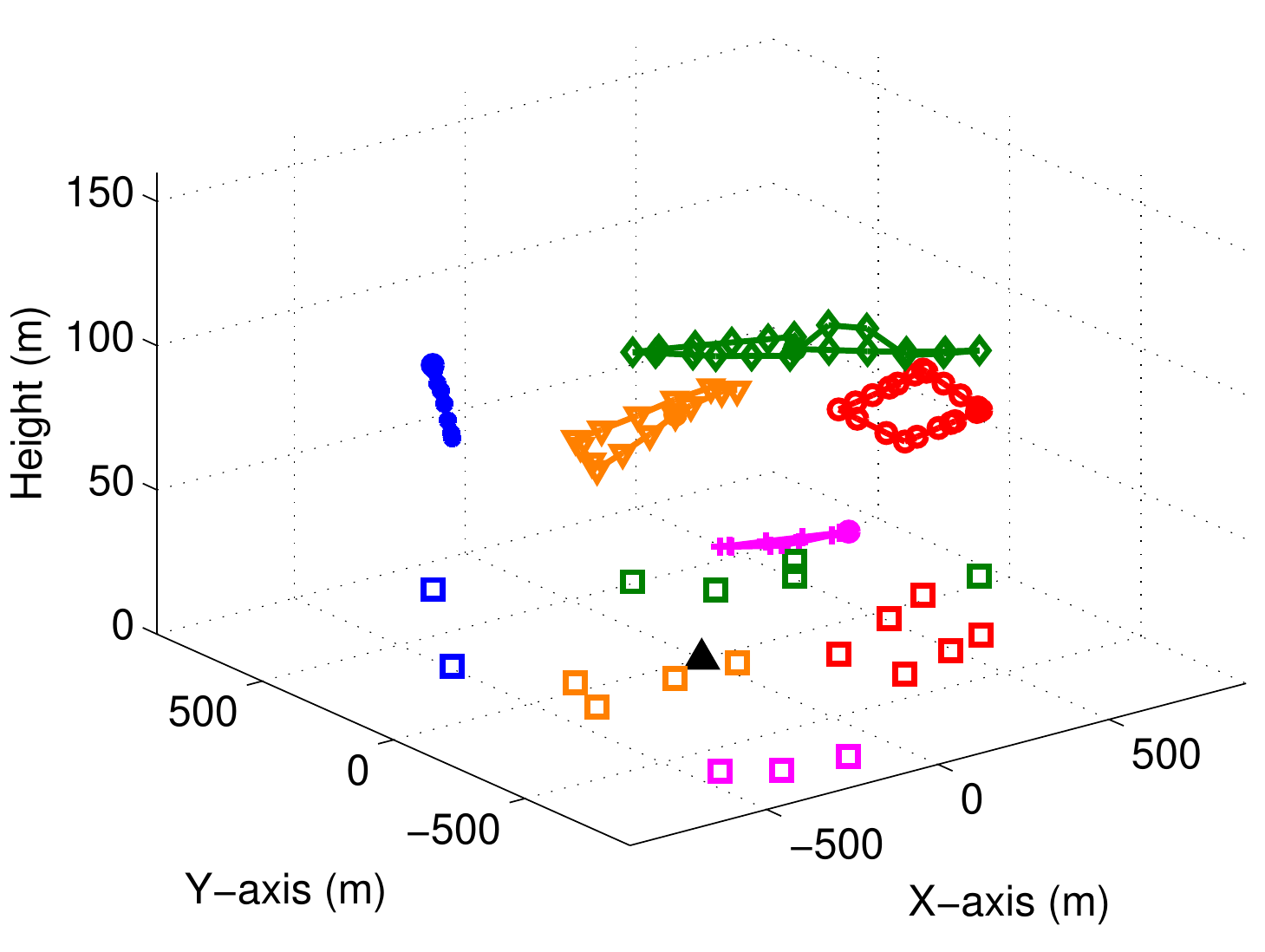}}
  \subfloat[Vertical view]{\includegraphics[width=0.33\textwidth]{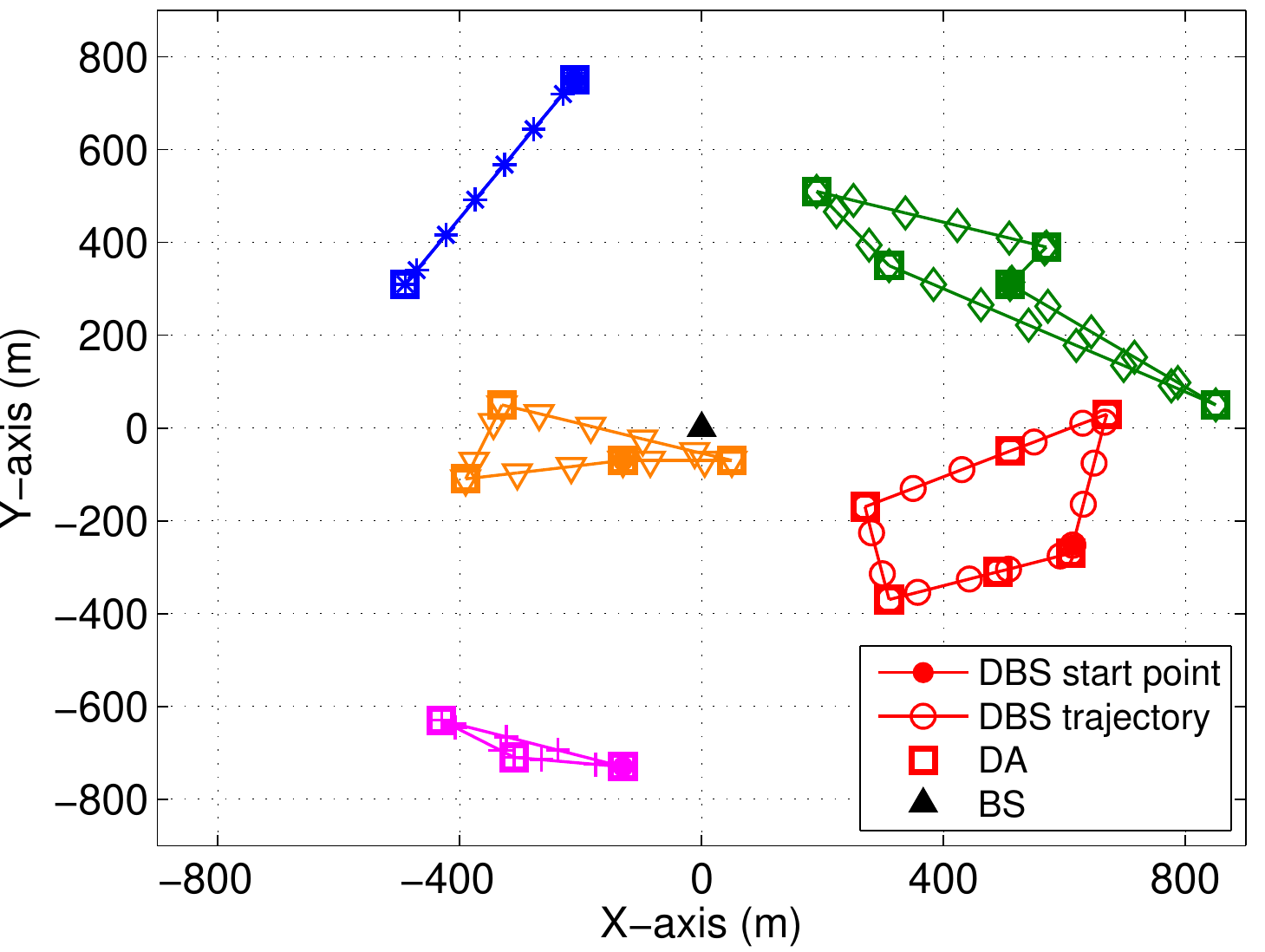}}
  \subfloat[Altitudes]{\includegraphics[width=0.33\textwidth]{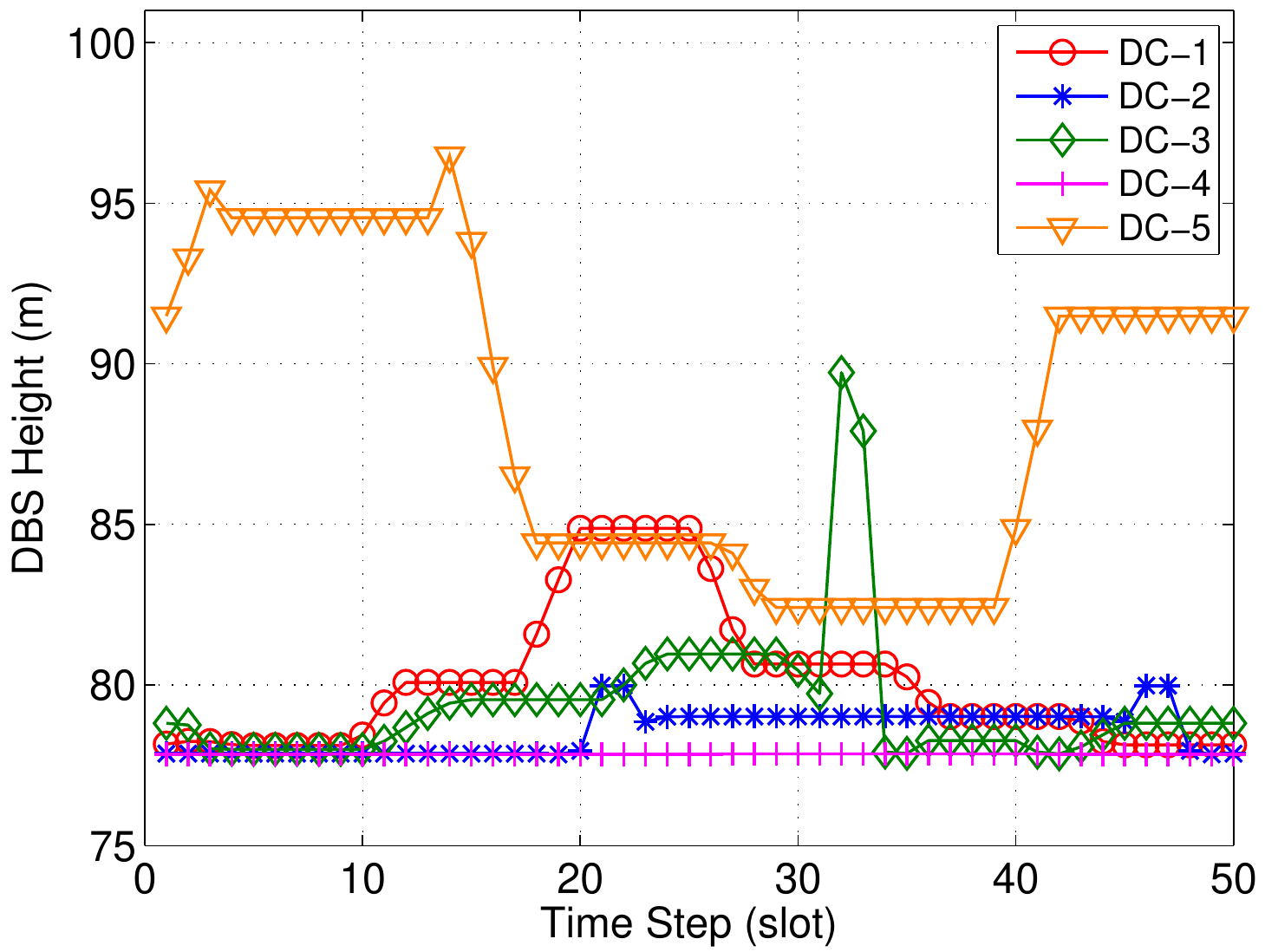}}
%   }
%   \subfigure[]{
%   \includegraphics[width=0.48\textwidth]{useUseTraj4_20_70z}
%   }
%   \subfigure[]{
%   \includegraphics[width=0.48\textwidth]{useUseHeight4_20_70}
%   }
%   \subfigure[]{
%   \includegraphics[width=0.48\textwidth]{useUseSchedule4_20_70}
%   }
  \caption{Trajectory design results of 5 DCs to cover 20 users.}
  \label{disFigall}
%   \vspace{-0.2cm}
\end{figure*}

\section{Simulation Results}
We implement the proposed algorithm in simulations by using Gurobi solver \cite{gurobi2018optimization}.
% In reality, the algorithm is executed by the BS which controls the flight of each DC to follow the optimized trajectory.
The parameters of both U2D and D2B pathloss models are set as suburban scenario \cite{al2017modeling}.
The carrier frequency for U2D communication $f_{c}$ is set as $2.4~\mathrm{GHz}$, which is widely supported by commercial drone products and prevents interference to the cellular system. 
% \cite{bisio2018unauthorized}.
D2B communications use the $850~\mathrm{MHz}$ LTE band \cite{al2017modeling}.
As the initial trajectories, DCs are uniformly deployed over the radio coverage area of terrestrial BS with the same flying altitude $90~\mathrm{m}$. 
Without loss of generality, we treat one slot $\delta_{t}$ as the minimal time unit to calculate related variables including $V_\mathrm{max} = 90~\mathrm{m}/\mathrm{slot}$, $H_\mathrm{max} = 10~\mathrm{m}/\mathrm{slot}$ and $T = N\delta_{t}$.
According to the general specifications of commercial drones whose maximal horizontal and ascent/descent speeds are $50-70~\mathrm{km}/\mathrm{h}$ and $3-6~\mathrm{m}/\mathrm{s}$ respectively, the approximate value of $\delta_{t}$ is around $5~\mathrm{s}$.
Table. \ref{Table_simulation} summarizes the detail simulation parameters. 
\begin{table}[htpb]
\centering
\caption{Simulation Parameters}
\label{Table_simulation}
\begin{tabular}{l|l}
\hline\noalign{\vskip 0.3mm}\hline
Parameter Name & Value\\
\hline
% \hline\noalign{\smallskip}
BS radio coverage radius $r_{\mathrm{BS}}$ & $900~\mathrm{m}$\\
User(IoT device) number $|\mathcal{U}|$ & $20$\\
U2D parameters $(\eta_{\mathrm{LoS}},\eta_{\mathrm{NLoS}},a,b)$ & (0.1,21,4.88,0.43)\\
D2B parameters $(\alpha,A,\theta_{0},B,\eta_{0})$ & (3.04,-23.29,-3.61,4.14,20.7)\\
Carrier frequencies (U2D, D2B) & $(2.4~\mathrm{GHz},~850~\mathrm{MHz})$\\
Duration of one period $T$ & $50~\mathrm{slots}$\\
D2B pathloss constraint $\mathrm{L}_{\mathrm{DB}}$ & $80~\mathrm{dB}$\\
Minimal number of slots $S_\mathrm{min}$ & $4$\\
Maximal horizontal speed $V_\mathrm{max}$ & $90~\mathrm{m}/\mathrm{slot}$\\
Maximal vertical speed $H_\mathrm{max}$ & $10~\mathrm{m}/\mathrm{slot}$\\
Trajectory difference $\mathbf{\epsilon}$ & $0.1~\mathrm{m}$ for each slot \\
\hline\noalign{\vskip 0.3mm}\hline
\end{tabular}
\end{table}

Fig. \ref{disFigall} shows an example of trajectory design result in which five DCs relay data from 20 IoT users. 
The closed curves with different markers in Figs. \ref{disFigall}(a) and \ref{disFigall}(b) denote the trajectories of different DCs, while the squares on the X-Y plane represent the users.
Users are associated to their corresponding DCs with same colors. 
Fig. \ref{disFigall}(c) shows the variations of flying altitude within one period, where the lower bound of all trajectories near $78~\mathrm{m}$ is constrained by the D2B link quality constraint.
As shown in Figs. \ref{disFigall}(a) and \ref{disFigall}(b), for each DC, the optimized trajectory can fly over all its associated user and form a closed trajectory in 3D space.
In Fig. \ref{disFigall}, the time interval between two adjacent dots on the same trajectory is exactly one slot. 
\begin{figure}[htbp]
  \centering
  \includegraphics[width=0.48\textwidth]{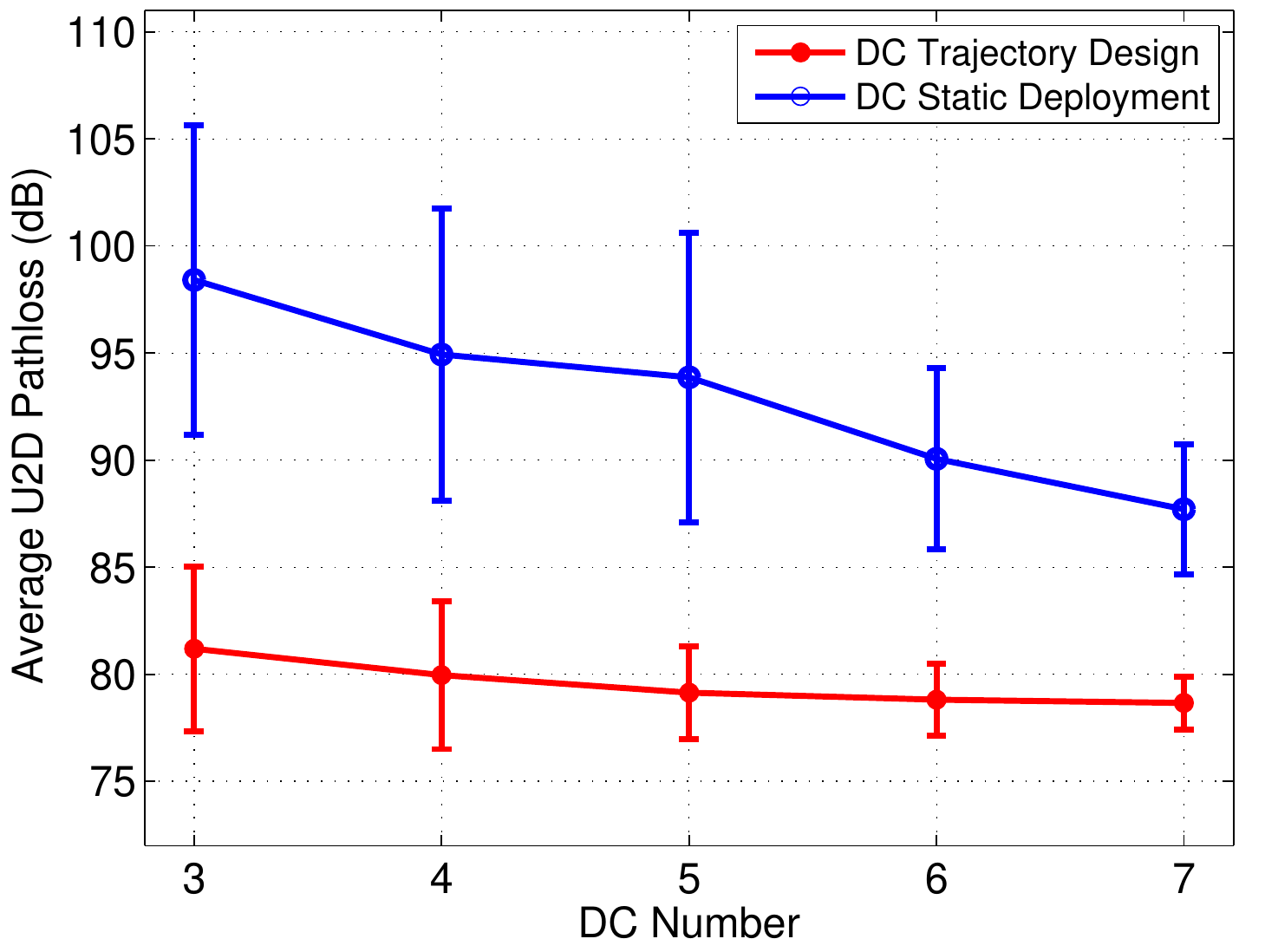}
  \caption{Average U2D pathloss comparison between trajectory design and static deployment.}
  \label{trajPsoCmp}
  \vspace{-0.5cm}
\end{figure}
In Fig. \ref{disFigall}(b), we can see that more than $50\%$ dots on each trajectory are overlapped above the associated users positions.
By joint analyzing Figs. \ref{disFigall}(b) and \ref{disFigall}(c), we can justify that DCs are prone to hovering above the associated users, while spending less slots for the travel process between two hovering positions. 
Such a ``hovering effect'' in the optimized trajectories can indicates the effectiveness of our proposed algorithm in minimizing average U2D pathloss.
Because with given flying altitude, the average U2D pathloss can be minimized when U2D horizontal distance is zero, i.e., hovering above the user.
% Therefore, the hovering effect in the trajectories can indicate the effectiveness of our proposed algorithm in minimizing average U2D pathloss.

\begin{figure}[htbp]
  \centering
  \includegraphics[width=0.48\textwidth]{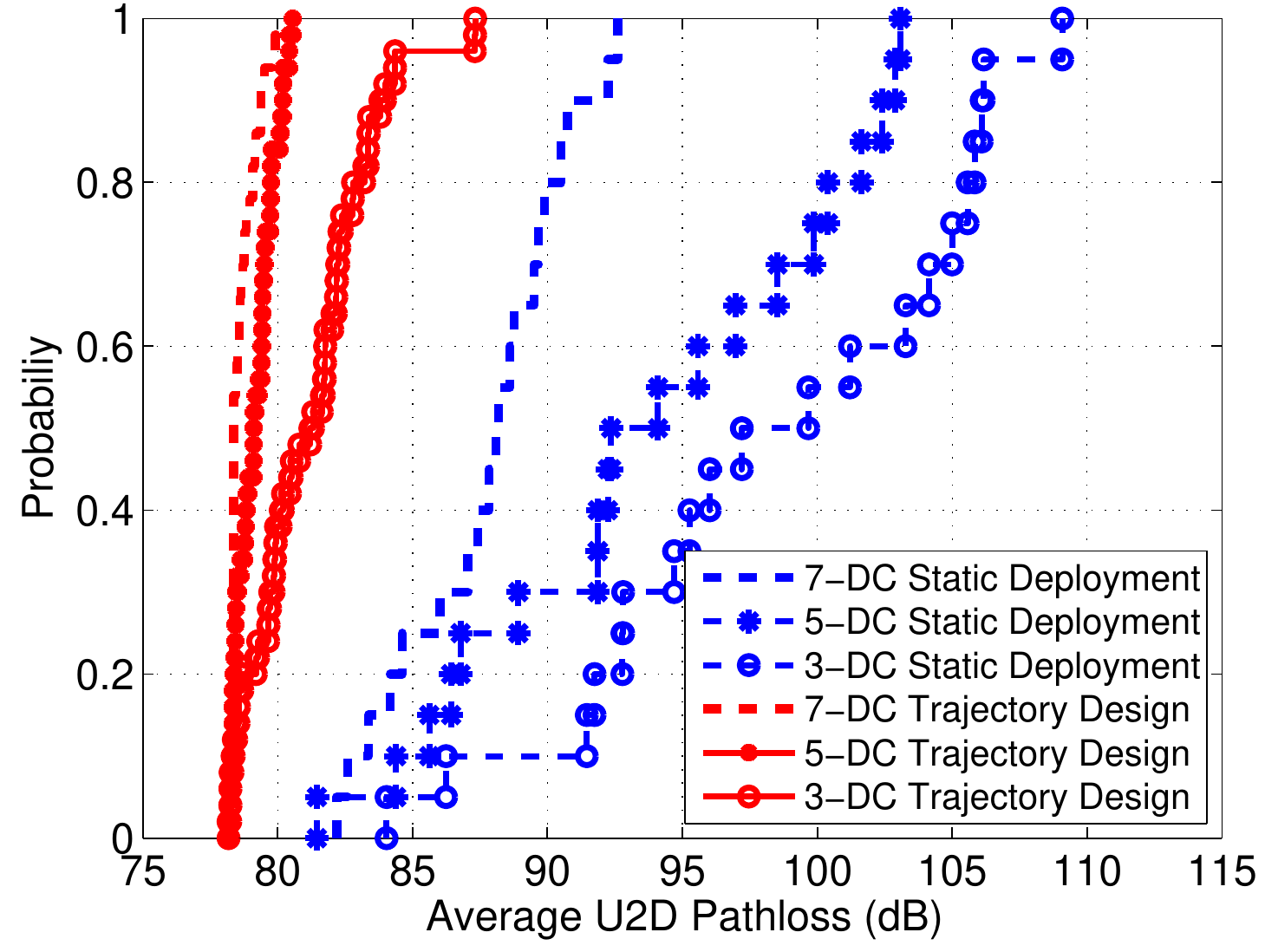}
  \caption{U2D pathloss CDF curves of trajectory design and static deployment.}
  \label{trajPerformCmp}
  \vspace{-0.5cm}
\end{figure}

Since most existing trajectory planning results are based on different D2G channel models with fixed-height (i.e. \cite{wu2017joint}), we choose one static DC deployment algorithm, the per-drone iterated particle swarm optimization (DI-PSO) algorithm \cite{shi2018multiple}, as the benchmark to highlight the efficiency of our proposed algorithm.
The average U2D pathloss performance achieved by the proposed 3D multi-DC trajectory design algorithm, as well as the static DC deployment scheme are compared in Fig. \ref{trajPsoCmp} and Fig. \ref{trajPerformCmp}.
In both figures, the blue curves represent static DC deployment while the red curves represent DC trajectory design.
From Fig. \ref{trajPsoCmp}, we can see that the U2D pathloss performance of both algorithms are improved as the available DC number increases. 
However, the average U2D pathloss of our trajectory design solution remains less than that of the DI-PSO solution by $10-15~\mathrm{dB}$.
The CDF curves in Fig. \ref{trajPerformCmp} further confirm that the trajectory design algorithm can maintain the U2D pathloss below given thresholds with higher probability when compared with static DC deployment. 
% , which indicates that the network performance of DA-RAN can be further improved through applying trajectory planning and scheduling approach to the statically deployed DCs.

The user fairness promotion provided by the DC trajectory design algorithm is indicated by the error bars in Fig. \ref{trajPerformCmp} and the U2D pathloss standard deviation comparison in Table \ref{Table_deviation}.
$\sigma_g$ and $\sigma_s$ are U2D pathloss standard deviations for DC trajectory design and static DC deployment, respectively.
In Fig. \ref{trajPerformCmp}, we can observe that the average U2D pathloss of static deployment ranges from $81~\mathrm{dB}$ to $119~\mathrm{dB}$, which is four times as much as that $(78-87~\mathrm{dB})$ of the trajectory design algorithm.
From Table \ref{Table_deviation}, we can conclude that the trajectory design can reduce the U2D pathloss standard deviation by $56.66\%$ on average compared with static deployment.
% the U2D pathloss standard deviation of DC trajectory design can be $20-60\%$ less than that of the static DC deployment.
% On average, the trajectory design can reduce the pathloss standard deviation by $44.62\%$ compared with static deployment.
\begin{table}[htpb]
\centering
\caption{U2D pathloss standard deviation comparison}
\label{Table_deviation}
\begin{tabular}{l|l|l|l|l|l}
\hline\noalign{\vskip 0.3mm}\hline
% \hline\noalign{\smallskip}
DC number & $3$ & $4$ & $5$ & $6$ & $7$ \\
\hline
$\sigma_g$ & $3.8481$ & $3.4554$ & $2.1818$ & $1.7013$ & $1.2311$ \\
\hline
$\sigma_s$ & $7.2211$ & $6.8313$ & $6.7562$ & $4.2329$ & $3.0530$ \\
\hline
$({\sigma_s} - {\sigma_g})/{\sigma_s}$ & $46.71\%$ & $49.42\%$ & $67.71\%$ & $59.81\%$ & $59.68\%$ \\
\hline\noalign{\vskip 0.3mm}\hline
\end{tabular}
\end{table}

\section{Conclusion}
% In this paper, we have proposed an efficient 3D trajectory design algorithm for multiple DCs to assist data collection in DC enabled IoT.
In this paper, we have investigated the 3D multi-DC trajectory design for efficient IoT data collection. 
An MINLP problem has been formulated to minimize the summation of average U2D pathloss.
% In this paper, we have formulated a MINLP problem which models the 3D trajectory design of multiple DCs to facilitate the IoT data collection. 
To solve the MINLP problem, we have decoupled it and formed multiple sub-problems in which the user association, U2D communication scheduling, horizontal trajectories, and flying altitudes are optimized, respectively.
% Then, the four sub-problems have been solved ineratively in the proposed 3D multi-DC trajectory design algorithm through BCD method.
Leveraging the BCD method, we have devised the 3D multi-DC trajectory design algorithm to solve the MINLP problem by solving sub-problems iteratively.
Simulation results have shown that the proposed DC trajectory design algorithm can achieve $10-15~\mathrm{dB}$ average U2D pathloss reduction, and promote pathloss standard deviation by more than $56\%$ when compared with the static DC deployment.
In future works, we will analyze the impacts of initial deployments, horizontal and vertical flying speeds, as well as inter-DC safe distance on the trajectory design, and investigate the communication and computation resources allocation of multiple DCs with the optimized trajectories.

\section*{Acknowledgment}
This work is supported by the National Natural Science Foundation of China under Project 91638204 and the Natural Sciences and Engineering Research Council (NSERC), Canada.

\ifCLASSOPTIONcaptionsoff
\newpage
\fi
%\newpage

\bibliography{referenceSWS}
\bibliographystyle{IEEEtran}

% that's all folks
% \end{spacing}
\end{document}